# Surface region band enhancement in noble gas adsorption assisted ARPES on kagome superconductor RbV$_3$Sb$_5$

Cao Peng,[1,*] Yiwei Li,[1,*] Xu Chen,[2] Shenghao Dai,[1] Zewen Wu,[3] Chunlong Wu,[1] Qiang Wan,[1] Keming Zhao,[1] Renzhe Li,[1] Shangkun Mo,[1] Dingkun Qin,[1] Shuming Yu,[1] Hao Zhong,[1] Shengjun Yuan,[3] Jiangang Guo,[2,4,‡] and Nan Xu[1,5,‡]

[1] *Institute for Advanced Studies, Wuhan University, Wuhan 430072, China*

[2] *Beijing National Laboratory for Condensed Matter Physics, Institute of Physics, Chinese Academy of Sciences, Beijing 100190, China*

[3] *School of Physics and Technology, Wuhan University, Wuhan 430072, China*

[4] *Songshan Lake Materials Laboratory, Dongguan 523808, China*

[5] *Wuhan Institute of Quantum Technology, Wuhan 430206, China*

* *These authors contribute equally.*

‡ *E-mails:* nxu@whu.edu.cn, JGGuo@iphy.ac.cn

**Electronic states near surface regions can be distinct from bulk states, which are paramount in understanding various physical phenomena occurring at surfaces and in applications in semiconductors, energy, and catalysis. Here, we report an abnormal surface region band enhancement effect in angle-resolved photoemission spectroscopy on kagome superconductor RbV$_3$Sb$_5$, by depositing noble gases with fine control. In contrast to conventional surface contamination, the intensity of surface region Sb band can be enhanced more than three times with noble gas adsorption. In the meantime, a hole-dope effect is observed for the enhanced surface region band, with other bands hardly changing. The doping effect is more pronounced with heavier noble gases. We propose that noble gas atoms selectively fill into alkali metal vacancy sites on the surface, which improves the surface condition, boosts surface region bands, and effectively dopes it with the Pauli repulsion mechanism. Our results provide a novel and reversible way to improve surface conditions and tune surface region bands by controlled surface noble gas deposition.**

# I. INTRODUCTION

Due to the sudden interruption of translation symmetry at crystal surfaces, electronic bands near the surface region could be quite different from the bulk states. Surface states have been experimentally discovered in various materials due to distinct origins and have contributed to multiple exotic surface properties, including nearly free two-dimensional electron gas (2DEG) with significant Rashba splitting on (111) surfaces of noble metals [1], emerging metallic 2DEG at surface/interface of insulating oxides [2-5] and robust surface states guaranteed by the bulk/boundary correspondence in a plethora of topological materials [6-10]. In semiconductors, surface region bands show bending behavior compared to bulk states, resulting from chemical potential shifting induced by surface potential. Surface band bending is one of the central concepts in semiconductor physics, which plays an important role in the semiconductor and photovoltaics industry [11].

Another case of discernible surface region band is materials with polar surfaces, in which different surface termination conditions effectively dope/modulate surface region band, in analog to surface potential induced band bending in semiconductors. One example is kagome material $AV_3Sb_5$ (A = K, Rb, Cs), which has drawn intensive attention recently due to unique band structures with flat bands, Dirac points, and van Hove singularities [12-19]. The natural cleavage plane is a polar surface between alkali metal (A) and Sb layers [Figs. 1(a)-1(b)]. Novel phenomena emerge at sample surfaces with different termination conditions [20-26]. The $p_z$ band from the Sb surface layer is more sensitive to surface conditions [27], which can be attributed to surface region bands. Consistent with previous results [28], Rb atoms on the polar surface evaporate gradually with time after sample cleavage, resulting in Sb-$p_z$ band near surface region significantly doped in $RbV_3Sb_5$ [Fig. 1(c)], while V-derived bands show negligible shift [Fig. 1(d)]. Therefore, investigations of the polar surface with A-vacancies and band structure near surface region are helpful to understand and manipulate these novel surface phenomena in $AV_3Sb_5$ material system.

In this work, we demonstrate that noble gas surface deposition is a novel way to boost and tune the Sb-$p_z$ surface region band in $RbV_3Sb_5$. By depositing Xenon (Xe) on the polar surface with fine control, the spectral intensity of the Sb surface region band gradually boosts up to more than three times, in contrast to regular surface adsorptions which diminish band intensity in ARPES. Furthermore, the Sb surface region band is hole-doped by approximately 60 meV. The doping effect is more pronounced with heavier noble atoms adsorbed. By removal of the surface adsorbed noble gases with an elevated temperature, the change on the Sb surface region band is fully reversible whereas other states are insensitive during the whole process. We attribute this abnormal spectral intensity increase and energy shift of the surface region state to the noble gas atom fillings on the alkali metal vacancy sites, which enhance surface conditions and dope the Sb surface region band by the Pauli repulsion mechanism. Our results could provide a novel technique to improve surface conditions and tune the surface region bands on polar surfaces with vacancies, which is helpful for both probes of surface region states using surface-sensitive experimental techniques and for further applications of surface region states.

## II. EXPERIMENTAL DETAILS

The high-quality single crystals of RbV$_3$Sb$_5$ were synthesized via the self-flux method [19]. Clean surfaces for ARPES measurements were obtained by cleaving samples *in situ* in a base vacuum better than $5\times10^{-11}$ Torr. The photoemission measurements were carried out at a lab-based ARPES system with a helium discharge lamp ($hv = 21.2$ eV). Noble gas adsorption on sample surfaces was achieved with partial pressure $\sim 1\times10^{-9}$ Torr at $T = 65$ K for Xenon.

## III. RESULTS

Fermi surface (FS) and band structure measured on the pristine surface of RbV$_3$Sb$_5$ [Figs. 2(a)-2(b)] are consistent with previous studies [29-33]. The $\alpha$ band derived from the Sb-$p_z$ orbit forms an electron-like pocket around the $\bar{\Gamma}$ point, while $\beta$ and $\gamma$ bands of V-orbit contribute to the van Hove singularities near Fermi energy ($E_F$) at the $\bar{M}$ point. Besides, another V-orbit-dominated band $\delta$ is observed as a triangular electron-like pocket around $\bar{K}$.

With Xe atoms adsorbed on the surface of RbV$_3$Sb$_5$, we discovered dramatic orbit-dependent effects on ARPES intensity. By carefully choosing deposition conditions (*e.g.*, sample temperature, Xe partial pressure, and deposition time), the intensity of the $\alpha$ band with Sb-orbit [Figs. 2(c)-2(d)] can be greatly enhanced compared with the

Page 3

results on the pristine surface [Figs. 2(a)-2(b)], whereas the intensity of V-orbit dominated bands $\beta$, $\gamma$ and $\delta$ slightly decrease. The distinction of intensity evolution between Sb- and V-orbits can be better visualized by the spectral weight difference of FS and band dispersion measurements before and after Xe adsorption, as shown in Figs. 2(e)-2(f), respectively.

Apart from the intensity change, we also observed an orbit-selective doping character induced by fine-tuned Xe adsorption. As the spectral intensity of the $\alpha$ band is boosted after Xe adsorption, its band bottom exhibits an upward energy shift of $\Delta E$ ~ 63 meV ($\Delta E = 63 \pm 2\ meV$) indicated by the comparison of energy-distribution curves (EDCs) at the $\overline{\Gamma}$ point [Fig. 2(g)]. Correspondingly, the Fermi momentum ($k_F$) of the $\alpha$ band shrinks by $\Delta k_F = 0.01$ Å$^{-1}$ ($\Delta k_F = 0.01 \pm 0.0004$ Å$^{-1}$), as extracted from the difference of the momentum-distribution curves (MDCs) at $E_F$ [Fig. 2(i)]. On the contrary, the doping level of V-orbit derived $\beta$ and $\gamma$ bands show negligible change with Xe adsorption. The overall band dispersion evolutions along the $\overline{\Gamma} - \overline{M}$ direction before and after Xe adsorption are summarized in Fig. 2(h), demonstrating a hole-doping effect selectively on the Sb-$p_z$ derived $\alpha$ band, with the amount estimated as 0.02 electron per unit cell. The $\alpha$ band shows a rigid band shift with effective mass change within 3%. Besides, the Xe adsorption causes no obvious effect on the CDW gaps of RbV$_3$Sb$_5$ (Fig. S1 in the Supplemental Material [34]). Our experimental observation is consistent with a previous STM study on RbV$_3$Sb$_5$ [26] showing an orbital-selective doping effect due to Rb desorption and a robust CDW gap regardless of different surface conditions.

To further understand the mechanism of the orbit-dependent band intensity and doping level modulations due to surface noble gas deposition, we record the fine time evolution of the spectrum along the $\overline{\Gamma} - \overline{M}$ direction during the whole Xe adsorption and desorption processes [Fig. 3(a)]. Compared to conventional monotonic intensity decreasing/increasing behaviors of $\beta$ band in the adsorption/desorption process, the $\alpha$ band shows exotic variations. With the temperature at 65 K in a base pressure of $5 \times 10^{-11}$ Torr (region-I in Figs. 3(b)-3(d)), the $\alpha$ band shows negligible changes within 30 minutes [Figs. 3(e-i)-3(e-ii)]. At the initial adsorption stage with a Xe partial pressure of $1 \times 10^{-9}$ Torr (region-II), the $\alpha$ band undergoes a sudden energy upward shift [Fig. 3(c)] accompanied by a dramatic intensity increase [Fig. 3(d)]. The effects induced by Xe adsorption saturates after ~ 40 minutes, with energy shift maxima of $\Delta E_1 = 67 \pm$ 4 $meV$ [Fig. 3(c)] and intensity boosting factor of I/I$_0$ ~ 3 [Fig. 3(d)]. By lowering sample temperature to $T$ = 55 K (region-III), the intensity drops rapidly [Fig. 3(d)] with stronger Xe adsorption and saturate after ~ 60 minutes, while doping level of the $\alpha$ band hardly changes [Fig. 3(c)].

These two-stage evolutions in region II and III can be perfectly reversed during the desorption process by increasing sample temperature (region IV and V in Figs. 3(b)-(d)), manifested by a constant doping level and uprising intensity (region-IV) followed by an electron-doping effect and descending intensity (region-V). The completion of the desorption process is marked by invariant doping level and intensity (region-VI), indicating a surface condition free of Xe atoms. EDCs at $\bar{\Gamma}$ extracted at typical stages of the adsorption and desorption processes [Fig. 3(e)] further demonstrate the unusual evolution of the $\alpha$ band.

From a microscopic view, we attribute the intensity increase of the $\alpha$ band in region-II to the filling of Xe atoms at the surface Rb vacancy sites (see the inset of region-II in Fig. 3(c)). Since the Rb vacancies break the long-range translation symmetry on the surface (inset of region-I in Fig. 3(c)), filling surface vacancies by Xe atoms can recover surface long-range order condition thus enhance the ARPES intensity. Specifically, the intensity reaches its maxima when all Rb vacancies are filled in by Xe atoms, which corresponds to the boundaries between region II/III (also see the EDC in Fig. 3(e-iii)). This scenario is further supported by the observation of non-dispersive Xe-5$p$ core levels in region-II (Fig. S2 in the Supplemental Material [34]) in great contrast to band dispersion of ordered Xe layers deposited on graphene forming high-order moiré pattern [35-37]. When the sample surface is below the critical temperature $T^*$ = 62 K in region III, more Xe atoms are adsorbed on the surface and form amorphous layers, leading to the decreasing intensity. By increasing sample temperature, the Xe amorphous layers first desorb gradually in region IV. In the temperature window of 62 K < $T$ < 69 K in region V, the improved surface with Rb vacancies occupied by Xe atoms is stable. Above 69 K, Xe atoms in the vacancy sites begin to detach gradually. The distinct time evolution behaviors between region-II (IV) and region-III (V) strongly indicate two well-separated phases of Xe adsorption (desorption) on the surface RbV$_3$Sb$_5$ and Rb vacancies being the energetically favorable occupation sites of Xe adatoms.

The observations of the nontrivial intensity boost and hole-dope effect of the surface Sb-$p_z$ derived state are reproducible for other noble gases including Ar and Kr, which implies a ubiquitous microscopic mechanism of noble gas atom fillings on the alkali vacancies, as illustrated in Figs. 4(a)-4(b). In addition, the saturated energy shift ($\Delta E$) is revealed to be proportional to the atomic mass of the adsorbed noble gas, as shown in Fig. 4(c) (also see Figs. S3-S4 in the Supplemental Material [34]), exhibiting a smaller $\Delta E$ below 30 meV for the lighter Ar atoms and a more significant $\Delta E$ about 60 meV for the heavier Xe atoms.

We propose that the Pauli repulsion between surface Sb-5$p_z$ derived states and the closed-shell noble gas atom is responsible for the hole-doping effect, as described in Fig. 4(d). The Sb atoms are exposed at Rb vacancy sites thus produce long-ranged evanescent charge densities into vacuum (see left panel of Fig. 4(d)). The electron clouds of the close-shell noble gas occupied on the Rb vacancy sites overlap with the Sb-5$p_z$ orbits and push electrons back into the bulk due to the Pauli exclusion principle (see right panel of Fig. 4(d)), leading to the observed hole-doping effect of the surface region Sb band. This interpretation is consistent with the atomic-mass-dependent energy shift [Fig. 4(c)] as the Pauli repulsion is strengthened in heavier noble gases due to larger electron shells and more pronounced orbit overlap [38,39].

Based on the charge conservation, the hole-doping of the surface region Sb-$p_z$ band would naturally cause the electron-doping of the bulk V-derived band. On the other hand, as the Pauli repulsion mechanism is a surface effect, the surface charges that are pushed into the bulk are greatly diluted. Therefore, the bulk V-derived states are hardly affected. The orbit-dependent doping effect induced by Pauli repulsion is confirmed by first-principle calculations (Fig. S5 in the Supplemental Material [34]).

## IV. CONCLUSIONS

In summary, we propose a novel and reversible method to improve polar surface conditions with surface alkali vacancies by noble gas adsorption, which leads to an intriguing surface region band spectral intensity enhancement and hole-doping effect. Our results not only provide an effective approach to desirable surface region band recovery with intriguing properties but also benefit surface-sensitive techniques like ARPES and scanning tunneling microscopy. In AV$_3$Sb$_5$, while no obvious effect is observed after Xenon adsorption on V-derived van Hove singularities [Fig. 2], Dirac points (see Fig. S6 in the Supplemental Material [34]) by ARPES, further investigations on this improved surface by STM may shed new light on understanding the novel surface phenomena that emerge at different surface conditions. Although V orbital derived bands of bulk origin play a significant role in the formation of CDW and superconductivity in the kagome superconductor family, we should not neglect the critical effect from the surface [20-26]. Notably, $CsV_3Sb_5$ has exhibited distinct properties of CDW and superconductivity on Cs- and Sb-terminated surfaces [21,24], which have not been fully understood. Therefore, our discovery can provide a novel technique for the study of CDW/superconductivity in the 135 kagome superconductor family under variant surface conditions.

This method might also be applied in other materials, which host polar cleaving surfaces with alkali metal or other kinds of vacancies, such as $K_xFe_2Se_2$ [40], $SmB_6$ [41,42], $Bi_2O_2Se$ [43] and possibly other nonlayered materials. It may also serve as a promising technique to improve the sample quality of two-dimensional thin films and devices, in which surface conditions can play a paramount role in their performances. Further, the reversible adsorption and desorption of noble gases via fine-controlled thermo-cycling promises applications based on surface states with precise doping level tunability.

**ACKNOWLEDGMENTS**

This work was supported by the National Natural Science Foundation of China (NSFC) (Grants No. 12274329, No. 52202342, and No. 12104304), the Ministry of Science and Technology of China (MOST) (Grants No. 2023YFA1406301), the Fundamental Research Funds for the Central Universities (Grants No. 2042023kf0107), and the China Postdoctoral Science Foundation (Grants No. 2023M732717).

# Figures

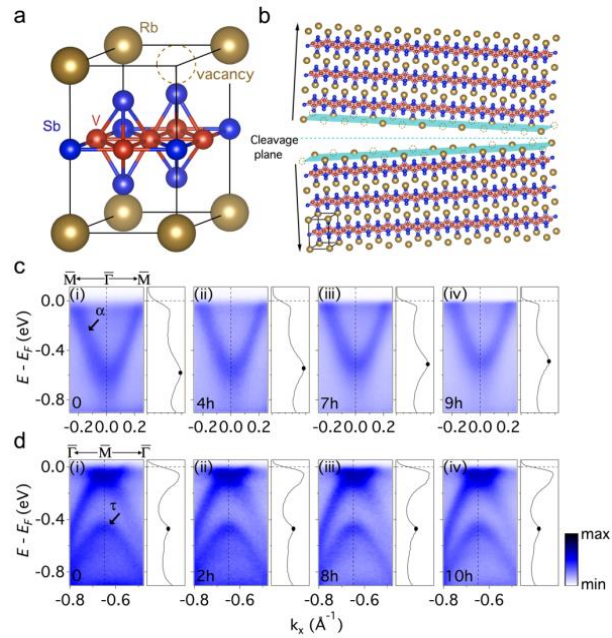

**FIG. 1. Polar cleaving surface and orbital-selective energy shift.** (a) Unit cell of RbV$_3$Sb$_5$. Rb, V, and Sb atoms are presented as gold, red, and blue balls, respectively. (b) Polar surface with Rb vacancies after fresh cleaving. Dotted circles denote Rb vacancies. (c),(d) Orbital-selective energy shift of Sb-$p_z$ derived $\alpha$ band and V-$d$ derived $\tau$ band with time.

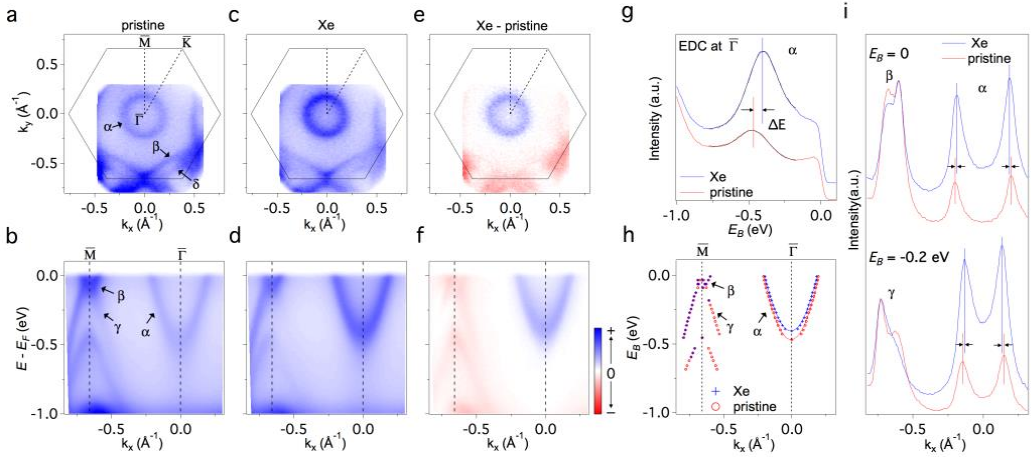

**FIG. 2. Surface region band enhancement and hole-dope effect induced by Xenon adsorption.** (a) Fermi surface with three pockets ($\alpha$, $\beta$, and $\delta$) measured on the pristine sample. (b) Band dispersion measured on the pristine sample along $\bar{\Gamma} - \bar{M}$ with three bands ($\alpha$, $\beta$, and $\gamma$). (c),(d) Fermi surface and band dispersion measured with Xe pressure ~ $1 \times 10^{-9}$ Torr at 65 K. (e),(f) Spectral intensity difference between the measurements on Xe-adsorbed and pristine samples (blue and red represent intensity increase and decrease, respectively). $\alpha$ pocket is greatly intensity-enhanced, while $\beta$, $\gamma$, and $\delta$ are intensity-reduced. (g) EDCs at $\bar{\Gamma}$ of the pristine (red) and Xe-adsorbed (blue) $\alpha$ pocket, showing clear energy upward shift ($\Delta E = 63 \pm 2\ meV$) as well as a strong coherence enhancement. Black lines show the results of single-peak Gaussian fit. (h) Traced energy bands obtained from (b) and (d), with red circles and blue plus represent pristine and Xe-adsorbed bands, respectively. $\alpha$ pocket shows a rigid band shift while $\beta$ and $\gamma$ remain still, indicating an orbital selective behavior. (i) MDCs at $E_B = 0$, -0.2 eV of the pristine (b) and Xe-adsorbed (d) cuts. Blue curves are normalized by peaks of $\beta$ and $\gamma$ bands. $\alpha$ pocket shrinks by $\Delta k_F = 0.01$ Å$^{-1}$ at $E_F$ and $\Delta k = 0.017$ Å$^{-1}$ at -0.2 eV, while $\beta$ and $\gamma$ band show no obvious change.

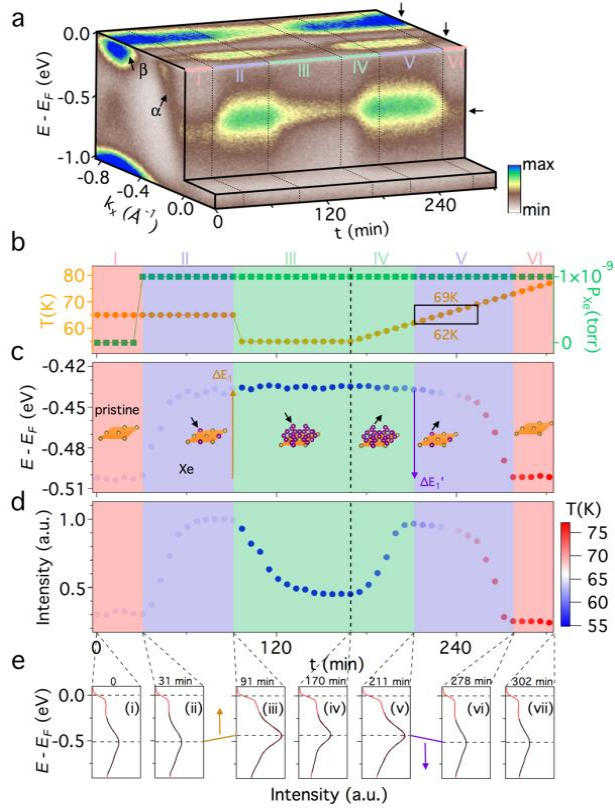

**FIG. 3. Elimination and recovery of surface vacancies by *in situ* Xenon adsorption and desorption.** (a) Spectral intensity evolution of the whole adsorption and desorption process. The spectrum intensity of $\beta$ band monotonically decreases and then increases, while $\alpha$ pocket shows an unconventional behavior. (b) Temperature (orange) and Xenon partial pressure (green) conditions. (c) Energy position of the band bottom of $\alpha$ pocket. (d) Intensity of the band bottom of $\alpha$ pocket. The distinct energy shifts and intensity evolution strongly indicate six well-separated phases indicated by I-VI. Insets in (c) show illustrations of different Xe adatom configurations on the sample surface. The color variation of data points in (c-d) indicates the temperature of the measurement. (e) Corresponding EDCs at $\bar{\Gamma}$ at typical stages of region I-VI. The Black box in region V in (b) denotes the temperature window (62 K < $T$ < 69 K) in which the improved surface with Xe atoms occupied on Rb vacancies is stable.

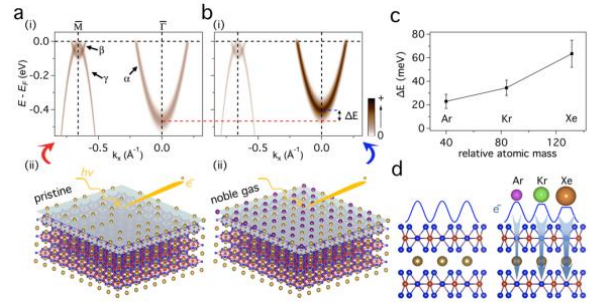

**FIG. 4. Pauli repulsion mechanism.** (a),(b) Schematics of spectra along $\bar{\Gamma} - \bar{M}$ measured on the pristine and Xe-adsorbed samples. The intensity of $\alpha$ band is greatly boosted and a hole-dope effect is observed. (c) Atomic-mass-dependent energy shift of $\alpha$ band induced by adsorbed noble gases (Ar, Kr, Xe). (d) Schematic illustration of Pauli repulsion mechanism. The Pauli exclusion causes the charge transfer from Sb surface region states to bulk states after noble gas adsorption.